\documentclass{article}
\usepackage{spconf,amsmath,graphicx}

\usepackage{amsmath,graphicx}
\usepackage{amssymb,amsmath,epsfig}
\usepackage{lipsum,color,multirow,url}
\usepackage{arydshln, algorithm, algpseudocode}
\usepackage{microtype}
\usepackage{nth}
\usepackage[inline]{enumitem}
\usepackage{romannum}
\usepackage{textcomp}
\usepackage{adjustbox}
\usepackage{bm}
\usepackage{comment}

\title{ Dual Application of Speech enhancement for automatic speech recognition}
\name{Ashutosh Pandey$^{1, \dagger}$, Chunxi Liu$^{2}$,  Yun Wang$^{2}$, Yatharth Saraf$^{2}$ 
\thanks{$\dagger$ Work was done when Ashutosh was an intern at Facebook.}}
\address{$^{1}$The Ohio State University, USA  \quad  $^{2}$Facebook AI, USA   \\
{\small \tt pandey.99@osu.edu \quad \{chunxiliu,yunwang,ysaraf\}@fb.com}  } 

\begin{document}
\ninept
\maketitle
\begin{abstract}
In this work, we exploit speech enhancement for improving a recurrent neural network transducer (RNN-T) based ASR system. We employ a dense convolutional recurrent network (DCRN) for complex spectral mapping based speech enhancement, and find it helpful for ASR in two ways:
a data augmentation technique, and a preprocessing frontend. In using it for ASR data augmentation, we exploit a KL divergence based consistency loss that is computed between the ASR outputs of original and enhanced utterances. In using speech enhancement as an effective ASR frontend, we propose a three-step training scheme based on model pretraining and feature selection. We evaluate our proposed techniques on a challenging social media English video dataset, and achieve an average relative improvement of 11.2\% with speech enhancement based data augmentation, 8.3\% with enhancement based preprocessing, and 13.4\% when combining both.
\end{abstract}
%
\section{Introduction}
\label{sec:intro}

Thus far we have seen a great deal of interest in automatic speech recognition (ASR) technologies, e.g., transcribing social media videos and enabling video captioning in both real-world high- and low-resource scenarios \cite{liao2013large, soltau2016neural, chiu2019comparison, liu2020multilingual, liu2020contextualizing}.    
End-to-end ASR models \cite{graves2012sequence, chan2016listen, sak2017recurrent} - that use a single neural network to transduce audio into word sequences - have been shown preferable 
in both (i) accuracy, i.e. as top performing models in various benchmarks \cite{chiu2019comparison, li2020comparison}, 
and (ii) training process which enables single-step learning from scratch. 
Thus in this work, we aim to exploit recurrent neural network transducer (RNN-T) \cite{graves2012sequence} for transcribing heterogeneous social media videos. 

A series of past efforts have been focused on improving RNN-T training, e.g., encoder and decoder pretraining \cite{rao2017exploring, hu2020exploring}, training optimization and criterion \cite{li2019improving, weng2019minimum},  etc. 
However, background noise and room reverberation in real world environment can still severely degrade ASR performance \cite{wang2018supervised, haeb2019speech}. In practical settings, the most popular approach to noise robust ASR is multi-channel speech enhancement that exploits spatial correlation between signals recorded using different microphones \cite{heymann2017beamnet, wang2020complex}. However, using multiple microphones is expensive and not always feasible. For example, social media videos are generally recorded using single microphones. 

Critical to the fact - why ASR systems based on single-channel speech enhancement can only obtain limited performance improvements - is mainly the observation that, processing artifacts introduced by single channel speech enhancement can dominate the benefits obtained by noise reduction \cite{jahn2016wide, wang2019bridging}. 
Recently, Wang et al. \cite{wang2020complex} proposed a complex spectral mapping based speech enhancement system that shows   significant ASR improvements via both single channel and multi-channel speech enhancement. Similarly, Kinoshita et al. \cite{kinoshita2020improving} have shown that using a time-domain speech enhancement can significantly improve ASR performance. Thus, presumably the recent success in applying speech enhancement to ASR is attributed to the joint enhancement of both magnitude and phase, which introduces less distortions compared to the magnitude-only enhancement systems.

Given the growing interest in RNN-T based ASR, to date little attempt has been made to increase its noise robustness. Specially, to our best knowledge, there is no pre-existing work that exploits single-channel speech enhancement for improving RNN-T based ASR. 
Thus in this work, we exploit a complex spectral mapping based speech enhancement system toward this end. We use a dense convolutional recurrent network (DCRN) as an enhancement model similar to \cite{wang2020complex}. 

First, we present that, by enhancing the ASR training data, speech enhancement can work effectively as an ASR training data augmentation technique. 
We also note that when applying speech synthesis to augmenting ASR training data in \cite{wang2020improving}, promoting consistent predictions in response to real and synthesized speech has been shown to provide substantial ASR performance gains.
Thus, similar to \cite{wang2020improving}, we also exploit a consistency criterion - an additional KL divergence loss between the RNN-T outputs in response to original and enhanced speech.

Secondly, we propose a multi (three)-step training scheme to combine DCRN and RNN-T, where speech enhancement plays a key role - as a preprocessing frontend - in facilitating ASR. 
First step is to learn an RNN-T ASR model on original utterances while DCRN on clean and noisy speech pairs. 
In step 2,  initialize ASR model with the RNN-T learned in 1st step, and perform RNN-T training on the enhanced utterances generated by the learned DCRN. 
Finally, in step 3, DCRN and RNN-T are jointly fine-tuned with the ASR RNN-T criterion.    

Note that the processing artifacts resulted from speech enhancement can make the enhanced speech suboptimal for subsequent ASR modeling. We thus examine making a selective use of both original and enhanced speech. 
We propose a trainable selection module that learns to interpolate original and enhanced features.
Specifically, the selection module is trained via ASR criterion to output probability scores in each time-frequency (T-F) bin, and the scores are used to provide a weighted combination of original and enhanced features.  
%

Lastly, to combine the proposed data augmentation and preprocessing, we show that the RNN-T trained from data augmentation can be further improved using steps 2 and 3 of the above three-step training scheme. 
Meanwhile, we continue to exploit the KL divergence based consistency criterion between the enhanced versions of (i) the original speech and (ii) the original speech mixed with noise. 

We evaluate our proposed methods on a challenging task of transcribing English social media videos. We achieve an average relative word error rate reduction (WERR) of 11.2\% when using speech enhancement for data augmentation, 8.3\% via preprocessing, and 13.4\% when combining the both. 

The rest of the paper is organized as follows. Section 2 describes the proposed techniques in this study. Experimental settings, results and comparisons are discussed in Section 3. Section 4 concludes the paper. 

%
\section{Modeling Approaches}
\label{sec:model}

The proposed system in this study has two components: RNN-T for ASR and DCRN for speech enhancement. In this section, we first briefly describe RNN-T based ASR and then complex spectral mapping based speech enhancement. Next, we provide a detailed description of the proposed DCRN and its dual application for ASR.

\subsection{RNN-T}
\label{ssec:model}

A speech utterance can be represented as an acoustic feature vector sequence $\bm{a} = (\bm{a}_{1} \ldots \bm{a}_{T}$), where $\bm{a}_{t} \in \mathbb{R}^{d}$ and $T$ is the number of frames in $\bm{a}$. 
Similarly, denoting a grapheme set or a wordpiece inventory as $\mathcal{Y}$, a transcription can be represented as a sequence $\bm{y} = (y_{1} \ldots y_{U})$ of length $U$, where $y_{u} \in \mathcal{Y}$. ASR can be formulated as a sequence-to-sequence problem where input sequence is $\bm{a}$, and output sequence is $\bm{y}$ corresponding to its transcription.
 We define $\bar{\mathcal{Y}}$ as $ \mathcal{Y} \cup \{ \emptyset \}$, where $\emptyset$ is the blank label.

%

An RNN-T model parameterizes the alignment probability   using an encoder network (i.e. transcription network in \cite{graves2012sequence}), a prediction network and a joint network. 
The encoder performs a mapping operation, denoted as $f^{\text{enc}}$,  which converts $\bm{a}$ into an another sequence of representations $\bm{h}^{\text{enc}} = (\bm{h}_{1}^{\text{enc}} \ldots \bm{h}^{\text{enc}}_{T})$:
\begin{equation}
 \bm{h}^{\text{enc}} = f^{\text{enc}}(\bm{a}) 
\end{equation}
\noindent 
A  prediction network $f^{\text{pred}}$, based on RNN or its variants, takes as input both its state vector and the previous non-blank output label $y_{u-1}$ predicted by the model, to produce the new representation $\bm{h}^{\text{pre}}_u$ as 
\begin{equation}
    \textbf{h}^{\text{pred}}_{1:u} = f^{\text{pred}}(y_{0:(u-1)})
\end{equation}
\noindent where $u$ is output label index and $y_0$ is the blank label.
The joint network $f^{\text{join}}$  is a feed-forward network that combines encoder output $\bm{h}^{\text{enc}}_t$ and prediction network output $\bm{h}^{\text{pre}}_{u}$ to compute logits $\bm{z}_{t,u}$ as 
\begin{equation}
    \bm{z}_{t,u} = f^{\text{join}}(\bm{h}^{\text{enc}}_t, \bm{h}^{\text{pred}}_u) 
\end{equation}
\begin{equation}
\begin{split} 
    P(y_u| \bm{x}_{1:t}, y_{1:(u-1)} )   = \text{Softmax}(\bm{z}_{t,u})
\end{split}
\label{eq:posterior_}
\end{equation}
\noindent such that the logits go through Softmax function and produce a posterior distribution of the next output label over $\bar{\mathcal{Y}}$. 
Finally, an RNN-T loss is computed using the negative log posterior  
\begin{equation}
    \mathcal{L}^{\text{\tiny{RNN-T}}} (\theta) = - \log  P(\bm{y} |  \bm{a} ) 
\end{equation}
\noindent where $\theta$ denotes the model parameters.  Note that the encoder is seen as an acoustic model, and the combination of prediction and joint network is interpreted as a decoder.

\begin{figure}[!t]
\centering
\includegraphics[width=0.65\columnwidth, keepaspectratio]{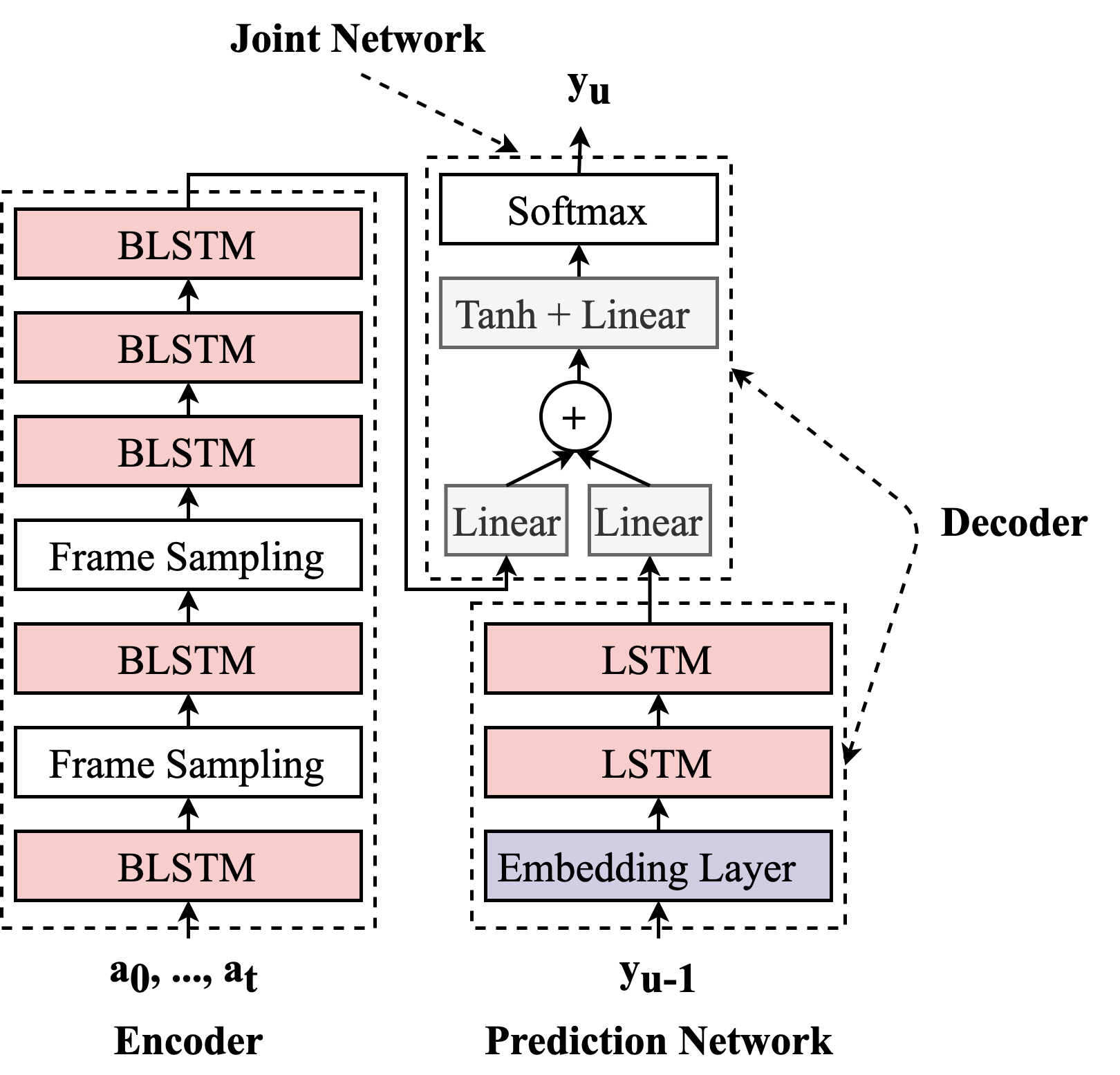}
\caption{\it The RNN-T ASR architecture used in this study.}
\label{fig:rnnt}
\end{figure}

Specifically in this work, we use an RNN-T architectures shown in Figure \ref{fig:rnnt}, which consists of a 5-layer BLSTM encoder and 2-layer LSTM decoder. The joint network combines the encoder and decoder outputs using linear layers as:  
\begin{equation}
\bm{z}_{t,u} = \text{Linear}(\text{Tanh}(\text{Linear}(\bm{h}_{t}^{\text{enc}}) +  \text{Linear}(\bm{h}_{u}^{\text{pred}})))
\end{equation}

\subsection{Speech Enhancement Using Complex Spectral Mapping}
Given a clean speech signal $\bm{s}$ and a noise signal $\bm{n}$, noisy speech signal $\bm{x}$ is defined as
\begin{equation}
    \bm{x} = \bm{s} + \bm{n}
\end{equation}

where $\{\bm{x}, \bm{s}, \bm{n} \} \in \mathbb{R}^{M \times 1}$, and $M$ is the number of samples. The objective of speech enhancement is to get a close estimate $\bm{\hat{s}}$ of $\bm{s}$ given $\bm{x}$. In short-time Fourier transform (STFT) domain, we get
\begin{equation}
\bm{X} = \bm{S} + \bm{N}
\end{equation}
where $\{\bm{X}, \bm{Y}, \bm{N} \} \in \mathbb{C}^{T \times F}$, $T$ is the number of frames and $F$ is the number of frequency bins. In Cartesian coordinates, a complex-valued STFT, such as $\bm{X}$, is represented as 
\begin{equation}
\bm{X} = \Re(\bm{X}) +i \cdot \Im(\bm{X})
\end{equation}

\noindent where $\Re(\bm{X})$ and  $\Im(\bm{X})$ respectively are the real and the imaginary components of $\bm{X}$. In complex spectral mapping, a real-valued DNN is employed to jointly predict the real and the imaginary components of $\bm{S}$ from the real and the imaginary components of $\bm{X}$.
\begin{equation}
\begin{split}
    [\Re(\bm{\hat{S}}), \Im(\bm{\hat{S}})] &= M_{\phi}([\Re(\bm{X}), \Im(\bm{X})])
\end{split}
\end{equation}
\noindent where $M_{\phi}$ denotes a speech enhancement model parameterized by $\phi$. Real and imaginary components are concatenated along the frequency dimension when processed using LSTMs or linear layers \cite{williamson2015complex, pandey2019exploring, pandey2020learning} and stacked to form the channel dimension when processed using convolutional layers \cite{fu2017complex, tan2019learning}. 

Finally, inverse STFT (ISTFT) is used to get the enhanced waveform.
\begin{equation}
\bm{\hat{s}} = \text{ISTFT}(\bm{\hat{S}}) = \text{ISTFT}(\Re(\bm{\hat{S}}) +i \cdot \Im(\bm{\hat{S}})) 
\end{equation} 

Next, we describe DCRN used in this study for complex spectral mapping.

\begin{figure*}[!t]
\centering\includegraphics[width=0.95\textwidth, keepaspectratio]{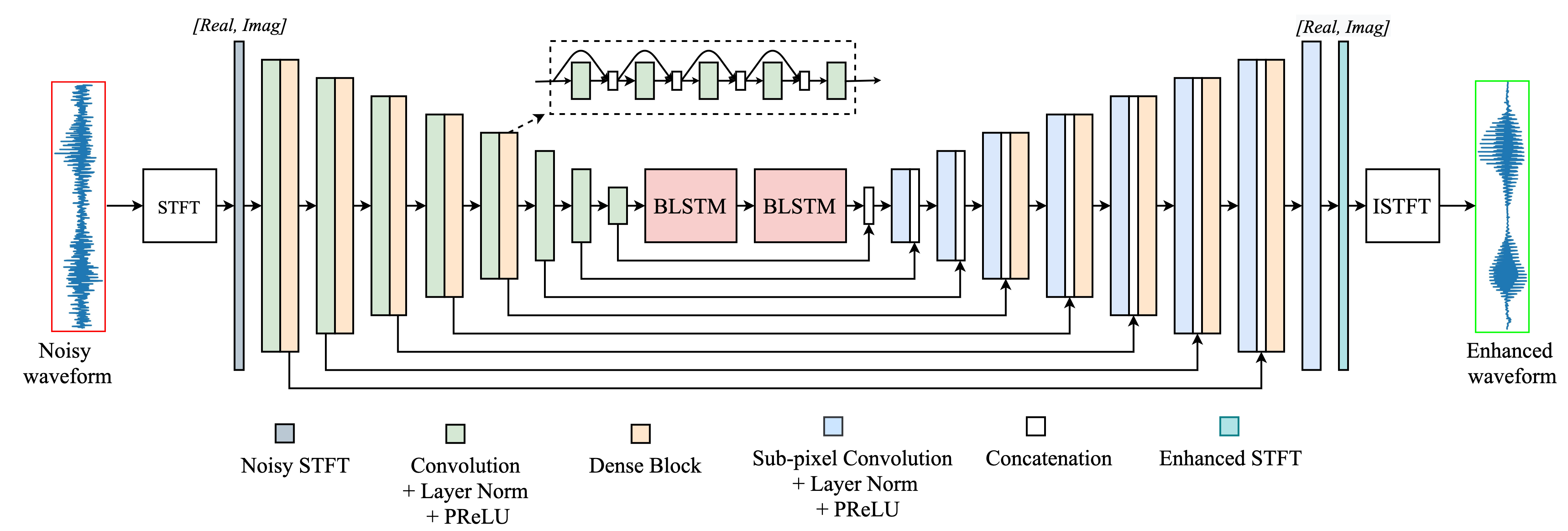}
\caption{\it The proposed DCRN model for complex spectral mapping.}
\end{figure*}


\subsection{Dense Convolutional Recurrent Network}

The proposed DCRN architecture is shown in Fig. 2. It is a 1D UNet architecture similar to \cite{wang2020complex}, which was proposed for robust ASR using single and multichannel speech enhancement. It consists of an encoder for downsampling, a decoder for upsampling, and two BLSTM layers between the encoder and the decoder for context aggregation over sequence of frames. Outputs from encoder layers are concatenated to the outputs from corresponding symmetric layers in decoder (along the channel axis). Downsampling in encoder is performed using convolutions with a stride of 2 along the frequency dimension, and upsampling in decoder is performed by using sub-pixel convolutions as in \cite{pandey2020densely}. 

Additionally, five layers in the encoder and five layers in the decoder are followed by a dense block. The dense block consists of five convolutional layers in which the input to a given layer is the concatenation of the outputs from all the previous layers in the block. The number of output channels after each convolution in a dense block is same as the number of channels in the input of dense block.

 The input to DCRN is $\bm{X}$ of shape $[BatchSize, 2, T, F]$ and output is $\hat{\bm{S}}$ of shape $[BatchSize, 2, T, F]$, where the real and the imaginary parts are stacked to form the channel axis. All the convolutions use filters of size $3 \times 3$ except at the input and the output, which use filters of size $5 \times 5$.  BLSTM layers use state size of 512 in both directions. The number of channels and the number of frequency bins in the outputs of the successive layers in the encoder and the decoder are $(2, 257)$ (input), $(32, 128)$, $(32, 64)$, $(32, 32)$, $(32, 16)$, $(64, 8)$, $(128, 4)$, $(256, 2)$, $(512, 1)$,  $(256, 2)$, $(128, 4)$, $(64, 8)$, $(32, 16)$, $(32, 32)$,  $(32, 64)$, $(32, 128)$, $(2, 257)$ (output).
 

\subsection{Dual Application of DCRN}
We explore speech enhancement using DCRN as a data augmentation technique and as a preprocessing frontend. Next, we describe these two techniques.
\subsubsection{Data augmentation}
Given an ASR data set $D$ consisting of speech signal and target sequence pairs $\{\bm{s}_{k}, \bm{y}_{k} \}$ where $k = 1, 2, \dots, N$, and $N$ is the number of data samples. We can obtain different versions of $\bm{s}_{k}$ for training in the following manner.
\begin{align}
    \bm{s}^{1} &= \bm{s} \\
    \bm{s}^{2} &= \bm{s} + \bm{n} \\
    \bm{s}^{3} &= \bm{\hat{s}} = M_{\phi}(\bm{s})\\
    \bm{s}^{4} &= \bm{\hat{s}}^{2} = M_{\phi}(\bm{s} + \bm{n})
\end{align}
where $\bm{n}$ is a noise signal added online during training, and we have dropped subscript $k$ for convenience. An ASR system can be trained on original speech $ \bm{s}^{1}$, speech with noise $\bm{s}^{2}$, enhanced speech $\bm{s}^{3}$, and enhanced version of speech with additive noise $\bm{s}^{4}$. Note that original speech in the social media dataset used in this study includes background noise, and we further add noise to it for noise based data augmentation.

\begin{figure}[!b]
\centering
\includegraphics[width=0.99\columnwidth, keepaspectratio]{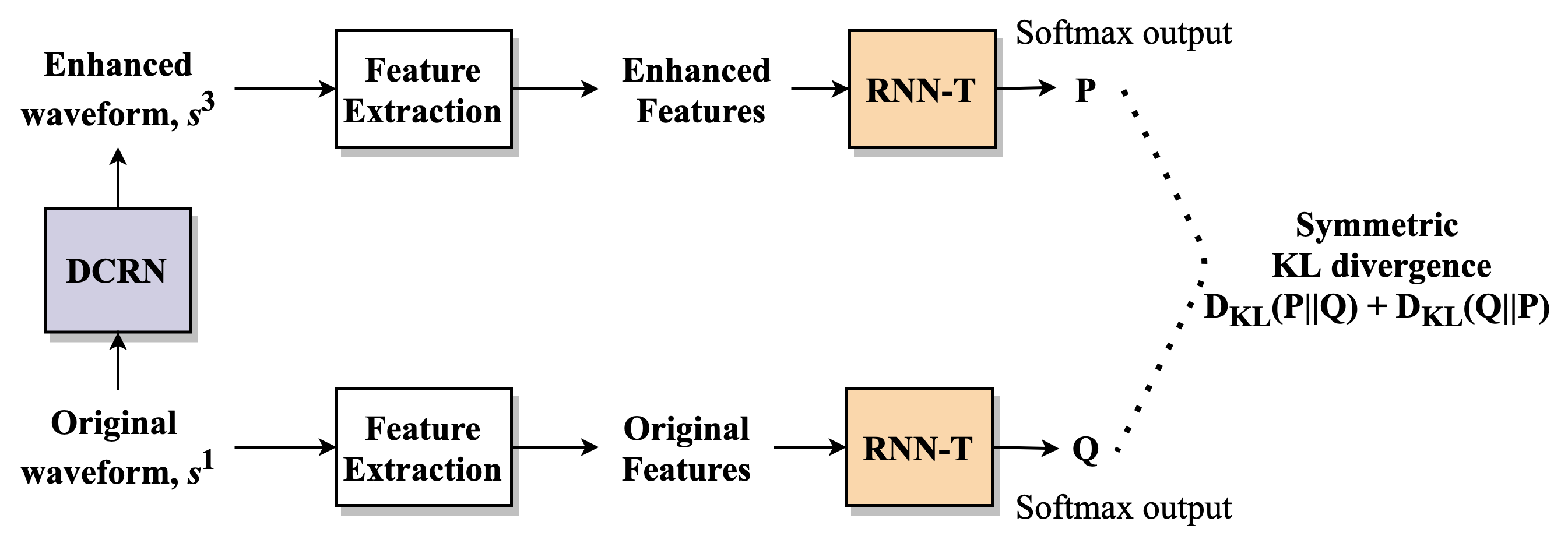}
\caption{\it An illustrative diagram of the KL divergence loss computation for pair $\{\bm{s}^{1}, \bm{s}^{3} \}$.}
\end{figure}

Further, since $\bm{s}^{1}, \bm{s}^{2}$,  $\bm{s}^{3}$, and $\bm{s}^{4}$ are different versions of a speech, we can use a KL divergence based consistency loss between their corresponding ASR outputs to make ASR model invariant to these variations. A similar idea of KL loss was used for speech synthesis based data augmentation in \cite{wang2020improving}. Given a pair $\{\bm{s}^{i}, \bm{s}^{j} \}$, we define KL divergence loss as following
\begin{equation}
\begin{split}
    \mathcal{L}^{\text{KL}}(\bm{s}^{i}, \bm{s}^{j}) = \frac{1}{T}\sum_{t=1}^{T}&[D_{\mathrm{KL}}(P(\bm{y}_{t}|\bm{s}^{i}) \| P(\bm{y}_{t}|\bm{s}^{j})) \\
    &+ D_{\mathrm{KL}}(P(\bm{y}_{t}|\bm{s}^{j}) \| P(\bm{y}_{t}|\bm{s}^{i}))]
 \end{split}
 \label{eq:kl}
\end{equation}
\noindent Note that RNN-T training computes each output posterior over both time index $t$ and label index $u$ as in Eq. \ref{eq:posterior_}, and we average the KL loss across label indices, while  omitting the label index $u$ in  Eq. \ref{eq:kl} for clarity. 

Thus the total training loss is defined as 
\begin{equation}
\begin{split}
    \mathcal{L}(\bm{s}^{i}, \bm{s}^{j}) = \ &0.5\cdot \log P(\bm{y}|\bm{s}^{i}) + 0.5\cdot \log P(\bm{y}|\bm{s}^{j}) \\ & + \lambda_{\mathrm{aux}}\mathcal{L}^{\text{KL}}(\bm{s}^{i}, \bm{s}^{j})
\end{split}
\end{equation}
where $\lambda_{\mathrm{aux}}$ is a hyperparameter determined using validation set. An illustrative diagram to compute KL loss between pair $ \bm{s}^{1}$ and  $\bm{s}^{3}$ is shown in Fig. 3. 

\subsubsection{Preprocessing}
For preprocessing, RNN-T is evaluated over test utterances enhanced using DCRN. We find that training RNN-T using enhanced utterance does not obtain better results. Instead, we train RNN-T with DCRN in three steps. First, RNN-T is trained using original utterances to get the baseline model, and DCRN is trained for speech enhancement using pairs of clean and noisy speech. Next, RNN-T is trained on enhanced utterances initialized using parameters from the baseline model. Finally, RNN-T and DCRN are jointly trained using a smaller learning rate. A schematic diagram of the three steps of RNN-T training with DCRN is given in Fig. 4.

\begin{figure}[!h]
\centering
\includegraphics[width=0.8\columnwidth, keepaspectratio]{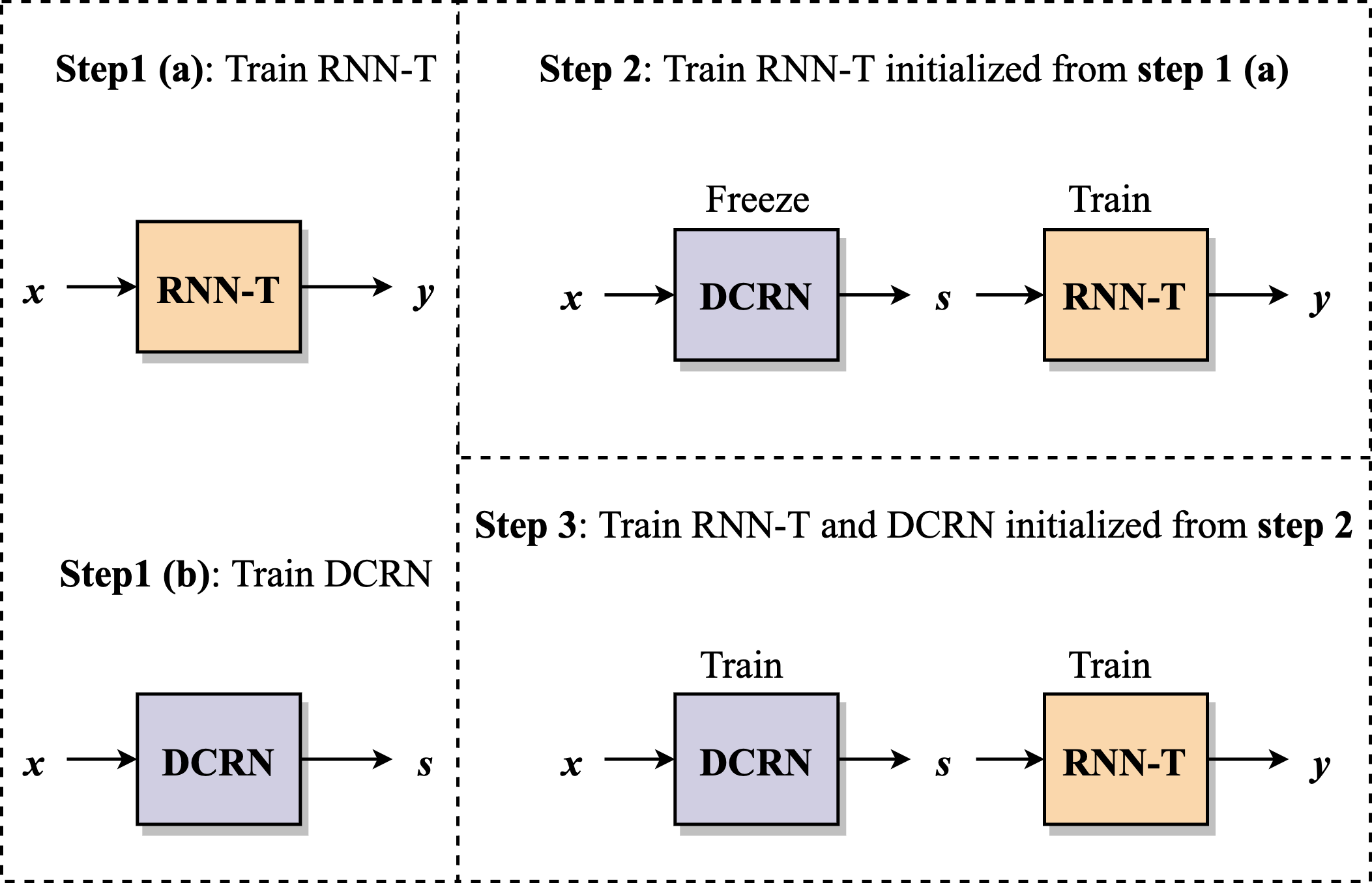}
\caption{\it Three-step training of RNN-T and DCRN.}
\end{figure}


\subsubsection{Selection Module}
Speech enhancement generally introduces processing artifacts that can outweigh the gain obtained from noise reduction. We propose a trainable selection module that can learn to select between enhanced and original features to improve ASR performance. A schematic diagram of selection module is shown in Fig. 5. It outputs a probability value, $p(t, f)$ for each time-frequency bin that represents the reliability of original features for ASR. We use $1-p(t,f)$ as the reliability score of enhanced features for ASR. The final acoustic feature, $\bm{\bar{a}}$, is computed as
\begin{equation}
\bar{a}(t, f) = p(t,f) \cdot a(t,f) + (1-p(t,f))\cdot \hat{a}(t,f)
\end{equation} 

Selection module uses a linear layer with 128 hidden units at the input, which is followed by 2 BLSTM layers with state size 128 in both directions, and a linear layer at the output. To train selection module, we utilize pre-trained DCRN and RNN-T. First, DCRN and RNN-T are fixed to only train selection module using ASR loss. Next, RNN-T and selection module are jointly trained with fixed DCRN using a smaller learning rate. 

 \begin{figure}[!b]
\centering
\includegraphics[width=0.99\columnwidth, keepaspectratio]{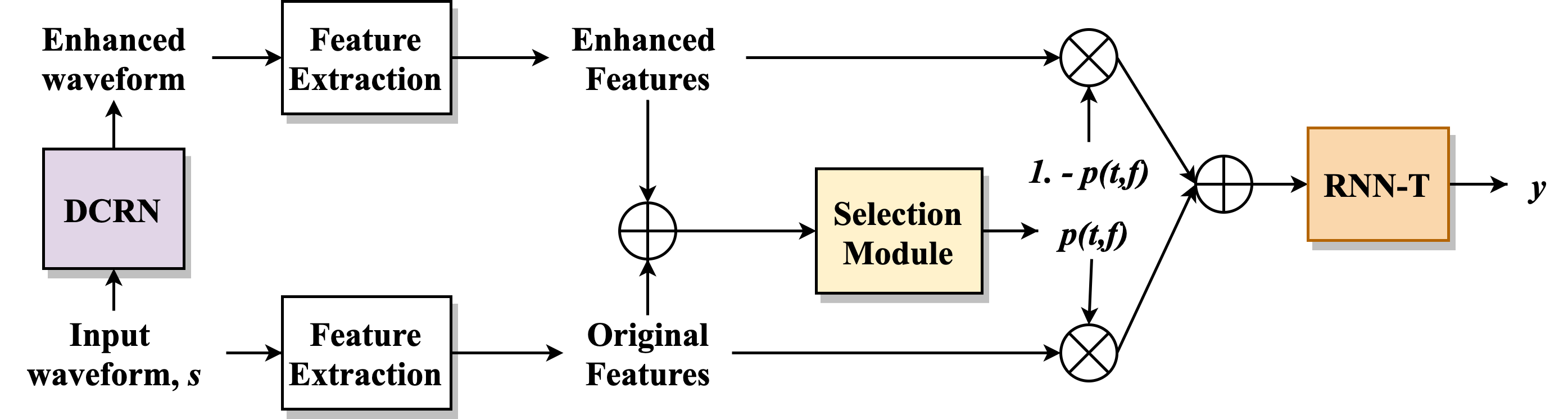}
\caption{\it The proposed selection module.}
\end{figure}

\section{Experiments}
\label{sec:exp}

\subsection{Data}
\label{ssec:data}

We evaluate our proposed methods on an in-house English video dataset.  The dataset is sampled  from  public  social  media  videos and de-identified before transcription.  
These videos contain a diverse range of acoustic conditions, speakers, accents and topics.  
The  test set contains two types of videos, \emph{clean} and \emph{noisy}. Train,  validation and test data sizes are given in Table 1.

\begin{table}[!h]
\caption{\label{tab:data} {\it Train, validation and test data duration in hours.}}
\centering 
\begin{adjustbox}{width=0.70\columnwidth}
\begin{tabular}{   c  c  c c c c  c  }
\hline \hline
        Language               &       Train  & Valid  &   \multicolumn{2}{   c  }{Test }  \\ 
                                &             &        &   clean         &     noisy \\
\hline \hline
\multirow{1}{*}{ }    English   &  2K       &      5.2     &    5.1 & 10.2\\
\hline \hline
\end{tabular}
\end{adjustbox}
\end{table}

Enhancement models are trained with a separate set of 1K hour clean audios, which are obtained also from the in-house English video dataset. 
Training utterances are generated using SNRs uniformly sampled from [-5 dB, 5 dB]. Enhancement models are evaluated at SNRs of -5 dB, 0 dB and 5 dB using 1000 utterances not used during training.

For generating noisy mixtures, we use non-speech classes from AudioSet  \cite{gemmeke2017audio}. For training RNN-T with additive noise based data augmentation, noises are randomly added to utterances with a probability of 0.5 at SNRs uniformly sampled from [0 dB, 25 dB]. For training RNN-T with enhancement based data augmentation, a given batch is randomly processed using DCRN with a probability of 0.5.

\subsection{Experimental Settings}
All speech utterances are resampled to 16 kHz and normalized to the range $[-1, 1]$. DCRN uses STFT with a frame size of 32 ms and frame shift of 10 ms. 
ASR features are extracted using 80 dimensional log Mel-filterbanks with a frame size of 16 ms and a frame shift of 10 ms. An utterance level mean and variance normalization is applied to Log-Mel features.

We apply the additional frequency and time masking  as in SpecAugment  \cite{park2019specaugment}. 
RNN-T output labels consist of a blank label and 255 word pieces generated by the unigram language model algorithm from SentencePiece toolkit \cite{kudo2018sentencepiece}.

All the models are trained using Adam optimizer \cite{kingma2014adam} using a tri-stage learning rate schedule proposed in \cite{park2019specaugment}. We use first 2 epochs for warm up and then divide remaining epochs in 3 equal parts. First part is trained with constant learning rate and the last 2 parts are trained using decaying learning rate, where minimum learning rate is  4e-6 and peak learning rate is 4e-4. Training in the third step of Fig. 4 uses a constant learning rate of 4e-6. 

RNN-T model used BLSTM encoder layers of  800 hidden units,  and a 2-layer LSTM decoder of 160 hidden units. 
Linear layers after encoder and decoder in the joint network use 1024 hidden units. Output after the first and second BLSTM layer in encoder are subsampled by a factor of two along the time dimension. The selection module uses two BLSTM layers of state size 128 and linear layers at input and output. $p(t, f)$ is computed using Sigmoidal nonlinearity at output.


\begin{table}[t]
\caption{\label{tab:data} {\it STOI and SI-SNR comparisons between DCRN and BLSTM.}}
\centering 
\begin{adjustbox}{width=0.90\columnwidth}
\begin{tabular}{|c|ccc|ccc|}
\cline{2-7}
\multicolumn{1}{c|}{} & \multicolumn{3}{c|}{ STOI (\%) } & \multicolumn{3}{c|}{ SI-SNR } \\
\hline
test SNR & -5 dB & 0 dB & 5 dB & -5 dB & 0 dB & 5 dB \\
\hline
original + noise & 57.3 & 68.6 & 78.9 & -5.0 & 0.0 & 5.0 \\
BLSTM & 75.1 & 82.6 & 87.4 & \textbf{6.6} & \textbf{9.8} & \textbf{12.1} \\
DCRN & \textbf{76.9} & \textbf{84.3} & \textbf{88.7} & 6.5 & 9.1 & 10.6 \\
\hline
\end{tabular}
\end{adjustbox}
\end{table}

\begin{table}[t]
\caption{\label{tab:data} {\it  WER and WERR comparisons between different data augmentation techniques on top of baseline. SE denotes speech enhancement and $L_{KL}$ denotes training with KL divergence loss.}}
\centering 
\begin{adjustbox}{width=0.95\columnwidth}
\begin{tabular}{|c|c|c|c|}
\cline{2-4}
\multicolumn{1}{c|}{}& Clean & Noisy & Average \\
\cline{1-3}
Baseline RNN-T & 14.8 & 19.4 & WERR \\
\hline
+ noise & 14.1 & 18.6 & 4.4\% \\
+ noise + $L_{KL}(\bm{s}^1, \bm{s}^2)$& 13.8 & 18.3 & 6.2\% \\
+ SE & 14.3 & 18.9 & 3.0\% \\
+ SE + $L_{KL}(\bm{s}^1, \bm{s}^3)$ & 13.9 & 18.4 & 5.6\% \\
+ noise + SE & 13.7 & 18.0 & 7.3\% \\
\hline
+ noise + SE + $L_{KL}(\bm{s}^{(1,2)}, \bm{s}^{(3,4)})$ & \textbf{13.0} & \textbf{17.4} & \textbf{11.2\%} \\
\hline
\end{tabular}
\end{adjustbox}
\end{table}

\begin{table}[t]
\caption{\label{tab:data} {\textit{WER comparisons between different training schemes. Model ids, such as a) and b) in the first column are used in the second column to denote the initialization model.} DCRN\_fixed \textit{denotes non-trainable DCRN.}}}
\centering 
\begin{adjustbox}{width=0.99\columnwidth}
\begin{tabular}{|c|c|c|c|c|c|}
\cline{3-6}
\multicolumn{1}{c}{}& \multicolumn{1}{c|}{}& \# params. & Clean & Noisy & Average \\
\hline
(a) & RNN-T & 90 M & 14.8 & 19.4 & WERR \\
\hline
& RNN-T + DCRN\_fixed & \multirow{4}{*}{ 103 M } & 15.0 & 19.9 & -2.0\% \\
(b) & (a) + DCRN\_fixed & & 14.1 & 18.8 & 3.9\% \\
(c) & (b) + fine tunning & & 13.5 & 18.0 & 8.0\% \\
& (c) + selection & & \textbf{13.4} & \textbf{18.0} & \textbf{8.3\%} \\
\hline
& RNN-T\_large & 105 M & 14.1 & 18.6 & 4.4\% \\
\hline
\end{tabular}
\end{adjustbox}
\end{table}

\begin{table}[t]
\caption{\label{tab:data} {\it WER  results in combining data augmentation and preprocessing.}}
\centering 
\begin{adjustbox}{width=0.9\columnwidth}
\begin{tabular}{|c|c|c|c|c|}
\cline{3-5}
\multicolumn{1}{c}{} & \multicolumn{1}{c|}{} & Clean & Noisy & Average \\
\cline{1-4}
& Baseline RNN-T & 14.8 & 19.4 & WERR \\
\hline
(a) & + noise + SE + $L_{KL}(\bm{s}^{(1,2)}, \bm{s}^{(3,4)})$ & 13.0 & 17.4 & 11.2\% \\
(b) & (a) + DCRN & 12.7 & 17.2 & 12.8\% \\
(c) & (b) + selection & 12.7 & 17.2  &  12.8\% \\  
(d) &  (a) + DCRN + $L_{KL}(\bm{s}^{3}, \bm{s}^{4})$ & \textbf{12.6} &    17.2  &  13.1\%  \\
\hline
(e) &  (a) + DCRN + selection + $L_{KL}(\bm{s}^{3}, \bm{s}^{4})$ & \textbf{12.6} & \textbf{17.1} & \textbf{13.4\%} \\
\hline
\end{tabular}
\end{adjustbox}
\end{table}


\subsection{Evaluation Metrics}
We compare enhancement models in terms of short-time objective intelligibility (STOI) \cite{taal2011algorithm} and scale-invariant signal-to-noise ratio (SI-SNR) \cite{le2019sdr}.

We evaluate ASR models  in terms of word error rate (WER).      %
In each comparison, we first compute the relative WER reduction (WERR) on \emph{clean}/\emph{noisy} as a percentage, and then take the unweighted average of two percentages, which we refer to as an average WERR.

\subsection{Results and Comparisons}
First, we compare DCRN with a BLSTM model which is shown in \cite{pandey2020learning} to improve cross-corpus generalization \cite{pandey2020cross} .  
As in Table 2,  DCRN is consistently better than BLSTM in terms of STOI for all SNR conditions, while BLSTM is better in terms of SI-SNR.
In this work,  we proceed with DCRN for ASR experiments. 

Next,  ASR results with data augmentation are shown in Table 3, where $L_{KL}(\bm{s}^{(1,2)}, \bm{s}^{(3,4)})$ denotes  training uniformly using either $L_{KL}(\bm{s}^1, \bm{s}^3)$ or $L_{KL}(\bm{s}^2, \bm{s}^4)$. 
 We observe that using additive noise as data augmentation can obtain better WER on both test sets with an average WERR of 4.4\%. 
 Next, training with an additional KL loss  as in Eq. \ref{eq:kl}  improves it  to 6.2\%.  
 
 Using speech enhancement as data augmentation obtains an average WERR of 3.0\%, and adding KL loss achieves 5.6\%. 
 Finally, using both noise and speech enhancement based data augmentation along with  $L_{KL}(\bm{s}^{(1,2)}, \bm{s}^{(3,4)})$ obtains the best  WERR  11.2\%. 
 We also experimented with KL loss between all possible pairs, but the results were similar. 

Next, we evaluate the ASR performance using DCRN as a preprocessing frontend. 
Results are given in Table 4 along with the respective total model parameters . 
First, we observe that training RNN-T from scratch on enhanced utterances degrades the performance. 
However, when we initialize RNN-T training with a model learned on original utterances (Step 2 in Fig. 4), we observe consistent improvements.
The average WERR in this case is 3.9\%. 
Additionally, when we fine tune DCRN with RNN-T using a small learning rate of 4e-6 (Step 3 in Fig. 4), average WERR improves to 8.0\%. 
Further, adding selection module slightly improves performance on clean test set and provides an average WERR of 8.3\%. 
Since RNN-T with DCRN has more parameters than the vanilla RNN-T baseline, we build a larger RNN-T model by increasing  encoder layers from 5 to 6, which has the total parameters comparable to DCRN + RNN-T. 
RNN-T\_large obtains 4.4\% WERR. 
Also, the number of parameters in selection module is negligible compared to RNN-T + DCRN, so we have reported the same number of parameters in these two cases.

Finally, we combine data augmentation and preprocessing techniques by initializing the RNN-T training with a model trained using data augmentation (last row in Table 3). We follow step 2 and step 3 of the three-step training scheme in this case, where the initialization model in step 2 is obtained from training with data augmentation. The results are given in Table 5. An average WERR of 12.8\% is observed in this case. Also, we can exploit KL loss in this case by using $L_{KL}(\bm{s}^{3}, \bm{s}^{4})$. Note that we can not use other types of pairs for KL loss because in this case RNN-T requires training on enhanced utterances. We obtain best WERR of 13.4\% by combining data augmentation with preprocessing, which includes selection module and uses $L_{KL}(\bm{s}^{3}, \bm{s}^{4})$ during training step 2 and step 3.

\section{Conclusions}

In this work, we have explored a complex spectral mapping based speech enhancement system to improve ASR performance. Our key finding is that using additive noise and speech enhancement as data augmentation - paired with a proposed KL divergence criterion between the ASR outputs of original and enhanced utterances - obtains significant performance improvements. Further, we have found that using a stepwise training of speech enhancement and ASR system can also improve WER substantially. Also, we have observed additional improvements  by  combining both speech enhancement based data augmentation and  preprocessing techniques.



\bibliographystyle{my_IEEEbib}
\bibliography{refs}

\begin{thebibliography}{10}

\bibitem{liao2013large}
H.~Liao, E.~McDermott, and A.~Senior,
\newblock ``Large scale deep neural network acoustic modeling with
  semi-supervised training data for {YouTube} video transcription,''
\newblock in {\em ASRU}, 2013.

\bibitem{soltau2016neural}
H.~Soltau, H.~Liao, and H.~Sak,
\newblock ``Neural speech recognizer: Acoustic-to-word {LSTM} model for large
  vocabulary speech recognition,''
\newblock {\em INTERSPEECH}, 2017.

\bibitem{chiu2019comparison}
C.-C. Chiu, W.~Han, Y.~Zhang, R.~Pang, S.~Kishchenko, P.~Nguyen, A.~Narayanan,
  H.~Liao, S.~Zhang, A.~Kannan, et~al.,
\newblock ``A comparison of end-to-end models for long-form speech
  recognition,''
\newblock in {\em ASRU}, 2019.

\bibitem{liu2020multilingual}
C.~Liu, Q.~Zhang, X.~Zhang, K.~Singh, Y.~Saraf, and G.~Zweig,
\newblock ``Multilingual graphemic hybrid {ASR} with massive data
  augmentation,''
\newblock in {\em Workshop on Spoken Language Technologies for Under-resourced
  languages and Collaboration and Computing for Under-Resourced Languages},
  2020.

\bibitem{liu2020contextualizing}
D.-R. Liu, C.~Liu, F.~Zhang, G.~Synnaeve, Y.~Saraf, and G.~Zweig,
\newblock ``Contextualizing {ASR} lattice rescoring with hybrid pointer network
  language model,''
\newblock in {\em INTERSPEECH}, 2020.

\bibitem{graves2012sequence}
A.~Graves,
\newblock ``Sequence transduction with recurrent neural networks,''
\newblock {\em arXiv preprint arXiv:1211.3711}, 2012.

\bibitem{chan2016listen}
W.~Chan, N.~Jaitly, Q.~Le, and O.~Vinyals,
\newblock ``Listen, attend and spell: A neural network for large vocabulary
  conversational speech recognition,''
\newblock in {\em ICASSP}, 2016.

\bibitem{sak2017recurrent}
H.~Sak, M.~Shannon, K.~Rao, and F.~Beaufays,
\newblock ``Recurrent neural aligner: An encoder-decoder neural network model
  for sequence to sequence mapping,''
\newblock in {\em INTERSPEECH}, 2017.

\bibitem{li2020comparison}
J.~Li, Y.~Wu, Y.~Gaur, C.~Wang, R.~Zhao, and S.~Liu,
\newblock ``On the comparison of popular end-to-end models for large scale
  speech recognition,''
\newblock {\em arXiv preprint arXiv:2005.14327}, 2020.

\bibitem{rao2017exploring}
K.~Rao, H.~Sak, and R.~Prabhavalkar,
\newblock ``Exploring architectures, data and units for streaming end-to-end
  speech recognition with rnn-transducer,''
\newblock in {\em ASRU}, 2017.

\bibitem{hu2020exploring}
H.~Hu, R.~Zhao, J.~Li, L.~Lu, and Y.~Gong,
\newblock ``Exploring pre-training with alignments for rnn transducer based
  end-to-end speech recognition,''
\newblock in {\em ICASSP}, 2020.

\bibitem{li2019improving}
J.~Li, R.~Zhao, H.~Hu, and Y.~Gong,
\newblock ``Improving {RNN} transducer modeling for end-to-end speech
  recognition,''
\newblock in {\em ASRU}, 2019.

\bibitem{weng2019minimum}
C.~Weng, C.~Yu, J.~Cui, C.~Zhang, and D.~Yu,
\newblock ``Minimum bayes risk training of {RNN}-transducer for end-to-end
  speech recognition,''
\newblock {\em arXiv preprint arXiv:1911.12487}, 2019.

\bibitem{wang2018supervised}
D.~Wang and J.~Chen,
\newblock ``Supervised speech separation based on deep learning: An overview,''
\newblock {\em IEEE/ACM Transactions on Audio, Speech, and Language
  Processing}, vol. 26, no. 10, pp. 1702--1726, 2018.

\bibitem{haeb2019speech}
R.~Haeb-Umbach, S.~Watanabe, T.~Nakatani, M.~Bacchiani, B.~Hoffmeister, M.~L.
  Seltzer, H.~Zen, and M.~Souden,
\newblock ``Speech processing for digital home assistants: Combining signal
  processing with deep-learning techniques,''
\newblock {\em IEEE Signal processing magazine}, vol. 36, no. 6, pp. 111--124,
  2019.

\bibitem{heymann2017beamnet}
J.~Heymann, L.~Drude, C.~Boeddeker, P.~Hanebrink, and R.~Haeb-Umbach,
\newblock ``Beamnet: End-to-end training of a beamformer-supported
  multi-channel asr system,''
\newblock in {\em ICASSP}, 2017, pp. 5325--5329.

\bibitem{wang2020complex}
Z.-Q. Wang, P.~Wang, and D.~Wang,
\newblock ``Complex spectral mapping for single-and multi-channel speech
  enhancement and robust {ASR},''
\newblock {\em IEEE/ACM Transactions on Audio, Speech, and Language
  Processing}, vol. 28, pp. 1778--1787, 2020.

\bibitem{jahn2016wide}
L.~D. Jahn~Heymann and R.~Haeb-Umbach,
\newblock ``Wide residual {BLSTM} network with discriminative speaker
  adaptation for robust speech recognition,''
\newblock in {\em Workshop on Speech Processing in Everyday Environments
  (CHiME’16)}, 2016, pp. 12--17.

\bibitem{wang2019bridging}
P.~Wang, K.~Tan, et~al.,
\newblock ``Bridging the gap between monaural speech enhancement and
  recognition with distortion-independent acoustic modeling,''
\newblock {\em IEEE/ACM Transactions on Audio, Speech, and Language
  Processing}, vol. 28, pp. 39--48, 2019.

\bibitem{kinoshita2020improving}
K.~Kinoshita, T.~Ochiai, M.~Delcroix, and T.~Nakatani,
\newblock ``Improving noise robust automatic speech recognition with
  single-channel time-domain enhancement network,''
\newblock in {\em ICASSP}, 2020, pp. 7009--7013.

\bibitem{wang2020improving}
G.~Wang, A.~Rosenberg, Z.~Chen, Y.~Zhang, B.~Ramabhadran, Y.~Wu, and P.~Moreno,
\newblock ``Improving speech recognition using consistent predictions on
  synthesized speech,''
\newblock in {\em ICASSP}, 2020.

\bibitem{williamson2015complex}
D.~S. Williamson, Y.~Wang, and D.~Wang,
\newblock ``Complex ratio masking for monaural speech separation,''
\newblock {\em IEEE/ACM Transactions on Audio, Speech, and Language
  Processing}, vol. 24, no. 3, pp. 483--492, 2015.

\bibitem{pandey2019exploring}
A.~Pandey and D.~Wang,
\newblock ``Exploring deep complex networks for complex spectrogram
  enhancement,''
\newblock in {\em ICASSP}, 2019, pp. 6885--6889.

\bibitem{pandey2020learning}
A.~Pandey and D.~Wang,
\newblock ``Learning complex spectral mapping for speech enhancement with
  improved cross-corpus generalization,''
\newblock in {\em INTERSPEECH}, 2020, pp. 4511--4515.

\bibitem{fu2017complex}
S.-W. Fu, T.-y. Hu, Y.~Tsao, and X.~Lu,
\newblock ``Complex spectrogram enhancement by convolutional neural network
  with multi-metrics learning,''
\newblock in {\em MLSP}, 2017, pp. 1--6.

\bibitem{tan2019learning}
K.~Tan and D.~Wang,
\newblock ``Learning complex spectral mapping with gated convolutional
  recurrent networks for monaural speech enhancement,''
\newblock {\em IEEE/ACM Transactions on Audio, Speech, and Language
  Processing}, vol. 28, pp. 380--390, 2019.

\bibitem{pandey2020densely}
A.~Pandey and D.~Wang,
\newblock ``Densely connected neural network with dilated convolutions for
  real-time speech enhancement in the time domain,''
\newblock in {\em ICASSP}, 2020, pp. 6629--6633.

\bibitem{gemmeke2017audio}
J.~F. Gemmeke, D.~P. Ellis, D.~Freedman, A.~Jansen, W.~Lawrence, R.~C. Moore,
  M.~Plakal, and M.~Ritter,
\newblock ``Audio {Set}: An ontology and human-labeled dataset for audio
  events,''
\newblock in {\em ICASSP}, 2017, pp. 776--780.

\bibitem{park2019specaugment}
D.~S. Park, W.~Chan, Y.~Zhang, C.-C. Chiu, B.~Zoph, E.~D. Cubuk, and Q.~V. Le,
\newblock ``{SpecAugment}: A simple data augmentation method for automatic
  speech recognition,''
\newblock {\em INTERSPEECH}, 2019.

\bibitem{kudo2018sentencepiece}
T.~Kudo and J.~Richardson,
\newblock ``Sentencepiece: A simple and language independent subword tokenizer
  and detokenizer for neural text processing,''
\newblock {\em arXiv preprint arXiv:1808.06226}, 2018.

\bibitem{kingma2014adam}
D.~P. Kingma and J.~Ba,
\newblock ``Adam: A method for stochastic optimization,''
\newblock in {\em ICLR}, 2015.

\bibitem{taal2011algorithm}
C.~H. Taal, R.~C. Hendriks, R.~Heusdens, and J.~Jensen,
\newblock ``An algorithm for intelligibility prediction of time--frequency
  weighted noisy speech,''
\newblock {\em IEEE Transactions on Audio, Speech, and Language Processing},
  vol. 19, no. 7, pp. 2125--2136, 2011.

\bibitem{le2019sdr}
J.~Le~Roux, S.~Wisdom, H.~Erdogan, and J.~R. Hershey,
\newblock ``{SDR}--half-baked or well done?,''
\newblock in {\em ICASSP}, 2019, pp. 626--630.

\bibitem{pandey2020cross}
A.~Pandey and D.~Wang,
\newblock ``On cross-corpus generalization of deep learning based speech
  enhancement,''
\newblock {\em IEEE/ACM Transactions on Audio, Speech, and Language
  Processing}, vol. 28, pp. 2489--2499, 2020.

\end{thebibliography}

\end{document}